\documentclass[12pt,final]{iopart}
\jl{4}
\usepackage{iopams}
\usepackage{a4dutch}
\usepackage{epsfig}

\begin{document}

\title{Statistical Mechanics of semi-classical colored Objects}
\author{M.~ Hofmann%
\footnote[3]{present address: Sun Microsystems GmbH, Langen, Germany}, M.~Bleicher%
\footnote[4]{Fellow of the Josef Buchmann-Foundation}, S.~Scherer, L.~Neise, H.~St\"ocker, W.~Greiner}
\address{Institut f\"ur
Theoretische Physik,  J.~W.~Goethe-Universit\"at,\\
D-60054 Frankfurt am Main, Germany}
\date{\today}
\begin{abstract}
A microscopic model of deconfined matter based 
on color inter\-actions between semi-classical quarks is studied. 
A hadronization mechanism is imposed to examine the properties 
and the disassembly of a thermalized quark plasma and to
investigate the possible existence of a phase transition
from quark matter to hadron matter.
\end{abstract}
\vspace{1cm}

The study of relativistic heavy-ion collisions is motivated
to a considerable extend by the search for and the unambiguous 
observation of a phase transition from confined, hadronic matter to
a deconfined state of QCD-matter dubbed the quark-gluon plasma
\cite{Sto86,McL86}. 

In the forthcoming experiments at RHIC (and later at LHC), 
the formation of a zone of quark-gluon plasma is generally expected.
The primary  stage of a collision at RHIC will be dominated 
by hard pQCD processes leading to the creation of a tremendous
number of quarks and gluons which are believed to form a zone of 
hot and dense and therefore expectedly deconfined partonic matter. 
This part of a heavy-ion collision has been described microscopically 
by partonic cascade models as VNI \cite{Gei92}. 
However, pQCD is, by definition, only applicable in reactions with large
momentum transfer $Q^2$. 
At SPS these partonic processes are strongly suppressed as compared to
hadronic interactions in the early stage. Here, the strong collective
motion of the impinging heavy nuclei may drive the system to
temperatures and densities beyond the hadronic level into a deconfined
phase. However, in both pictures, partonic or hadronic, the major part 
of particle production takes place in primary collisions 
within the first few fm/$c$ when the system is 
strongly compressed and heated. 

Most recently, a combination of partonic and hadronic cascades
has been established by connecting the VNI model with the UrQMD
model which finally copes with the hadronic secondary interactions
\cite{Bas99}.  

Unfortunately, a possible quark-gluon plasma phase dominated 
by \emph{soft}, non-perturbative QCD processes which mediates between parton
and hadron mode and intrinsically performs the hadronization process
is not dynamically treated.  The 
non-perturbative properties of QCD, which are crucial for 
this transition, impede the applicability of all common approaches to
a first-principle description of hadronization. Effective models have
to be constructed which allow a numerical calculation of observables
by simulating  the essential features of non-perturbative QCD. In
\cite{Reh98}, a dynamical approach based on the Nambu-Jona-Lasinio
model has been presented, in which quarks are propagated on 
classical trajectories while their effective masses are 
calculated self-consistently according to the NJL equations of motion. 
Hadron production is driven by $qq$ and $qh$ collisions. 
Unfortunately, this approach does not provide confinement 
and therefore is not suitable for the investigation of heavy-ion collisions. 
On the footing of the Friedberg-Lee Lagrangian, a similar study has
been performed in the chromodielectric model \cite{Tra98} completely
respecting confinement. Hadronization is performed by mapping
quark-gluon states onto irreducible representations of color
$SU(3)$. However, this method is numerically extremely expensive.
This prohibits the simulation of heavy-ion collisions. 

In this paper we present a semi-classical model which mimics the
properties of non-abelian QCD by the means of a two-body color
potential between quarks. In
addition, a dynamical hadronization criterion is defined which allows for
the consecutive migration from quark to hadronic degrees of freedom. 
The long term objective of this investigation is the unification of
the different species of microscopic models, partonic in the initial,
hadronic in the final stage of the reaction, into one single model,
which finally will allow the simulation of a complete heavy-ion
collision including a QGP phase transition. In this paper we shall
elaborate the major thermodynamic properties of the so-defined system 
which will justify the crude approximation by its phenomenological
implications. In a subsequent publication we will investigate the dynamical
evolution of the model and adopt it to more realistic initial conditions 
which then will allow to describe heavy-ion collisions.

\section{The model Hamiltonian}

The colored and flavored quarks are treated as semi-classical particles
interacting via a Cornell potential with color matrices \cite{Isgur:1985bm}. 
This interaction provides an effective description of the 
non-perturbative, soft gluonic part of QCD. The 
Hamiltonian reads
\[
{\mathcal{H}} = \sum_{i=1}^N\sqrt{p_i^2+m_i^2}+\frac{1}{2}\sum_{i,j}C_{ij}V(\left\vert\mathbf{r}_i-\mathbf{r}_j\right\vert)
\]
where $N$ is the number of quarks.
Four quark flavors ($u,d,s,c$) with current masses
$m_u=m_d=10$~MeV, $m_s=150$~MeV and $m_c=1.5$~MeV
are considered. The confining properties of $V(r)$ are ensured 
by a linear increase at large distances~$r$. At short distances, 
the strong coupling constant
$\alpha_s$ becomes small, yielding a Coulomb-type behavior as in QED.
This color Coulomb potential plus the confining part is
the well known Cornell-potential \cite{EICH75}
\[
V(r) = -\frac{3}{4}\frac{\alpha_s}{r}+\kappa\,r\;,
\]
which has successfully been applied to meson spectroscopy.
For infinite quark masses this inter-quark potential
has also been found in lattice calculations over a wide
range of quark distances \cite{BORN94}.
For small quark masses, retardation and
chromomagnetic effects should be included.
This is neglected in the present work. However, the linear
behavior at large distances seems to be supported by the success of
the string model even for zero quark masses
\cite{ANDER83}. 

The color matrix elements $C_{ij}$ regulate the sign and
relative strength of the interaction between two quarks/antiquarks,
respectively, depending on the color combination of the pair. 
The matrix $C_{ij}$ in the short range color 
interaction potential between quarks,
$V_{\rm color} = - C_{ij}\frac{3}{4}\frac{\alpha}{r}$,
can be calculated from the quark-gluon interaction part of the
QCD Lagrangian
\[
\mathcal{L}_\mathrm{int} =
\frac{g}{2}\,\bar{\Psi}\lambda_a\gamma_\mu\Psi G^\mu_a\;
\]
on the one-gluon exchange level represented by the Feynman diagram

\centerline{\epsfig{figure=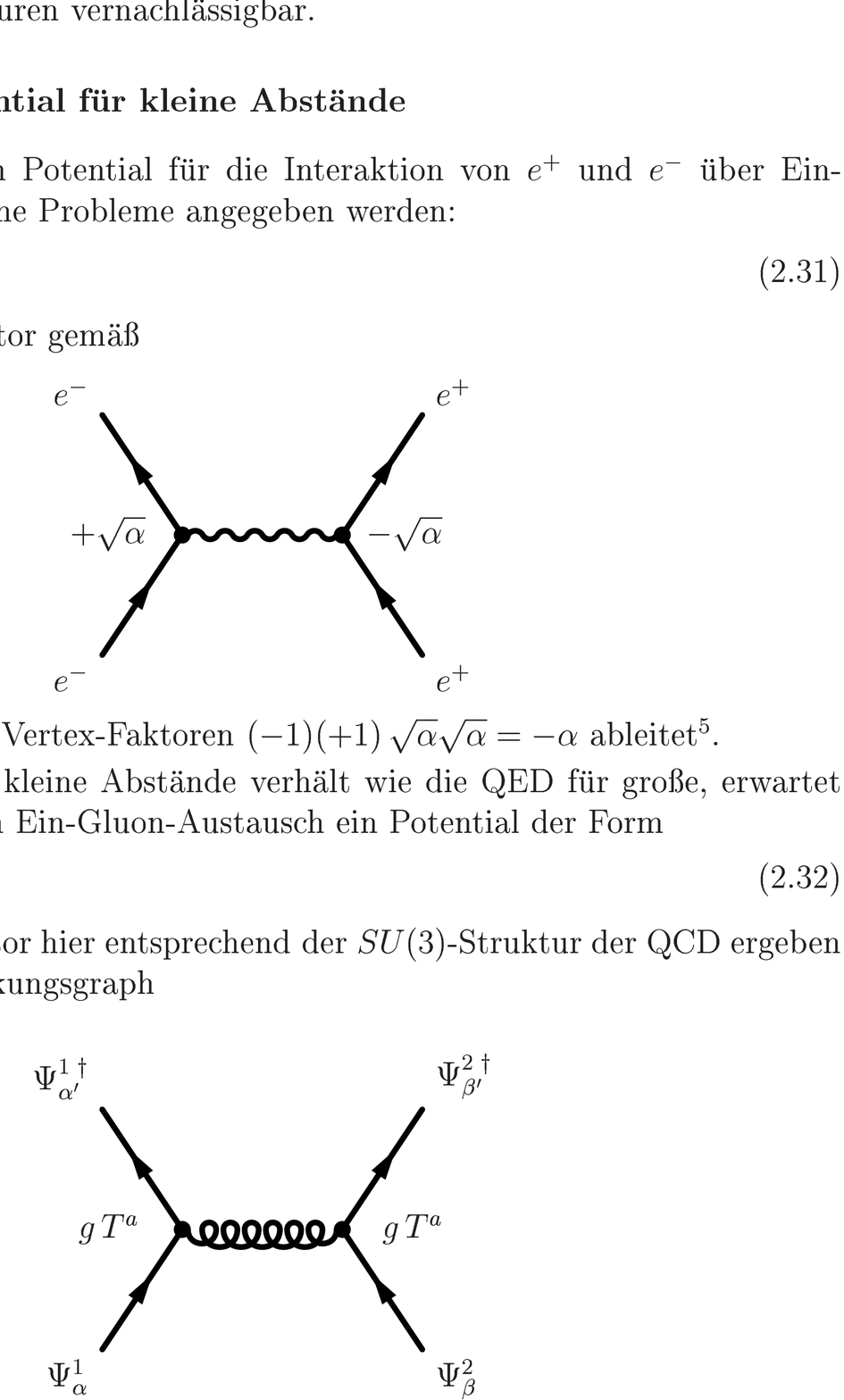,width=5cm,clip}}

Using the standard fundamental representation of $SU(3)_{\rm color}$ for the
quarks and the adjoint representation for the gluons,

\[
\vec{q}_R = \left(\begin{array}{c}
1\\ 0\\ 0
\end{array}\right),\quad
\vec{q}_G = \left(\begin{array}{c}
0\\ 1\\ 0
\end{array}\right),\quad
\vec{q}_B = \left(\begin{array}{c}
0\\ 0\\ 1
\end{array}\right),\quad
\]
\[
T^a = \frac{1}{2}\lambda^a,\quad a=1,\dots,8
\]
where $\lambda_a$ are the Gell-Mann matrices, and separating 
the quark wave function in the color and Dirac parts, 
\[
\Psi_\alpha = \psi\,\vec{q}_\alpha
\]
the interaction amplitude 
\[
\mathcal{M}_{\alpha\alpha^\prime\beta\beta^\prime}
\sim
\frac{g^2}{4}
\bar\Psi_{\alpha^\prime}\gamma_\mu\lambda^a\Psi_\alpha\,
\mathcal{D}^{\mu\nu}_{ab}(q)\,
\bar\Psi_{\beta^\prime}\gamma_\nu\lambda^b\Psi_\beta
\]
separates in color and Dirac parts ($\mathcal{D}^{\mu\nu}_{ab}(q) = D^{\mu\nu}(q)\delta_{ab}$ 
is the gluon propagator):
\[
\mathcal{M}_{\alpha\alpha^\prime\beta\beta^\prime}
\sim
\bar\psi_1\gamma_\mu\psi_1D^{\mu\nu}(q)\psi_2\gamma_\nu\psi_2\,
\vec{q}_{\alpha^\prime}^{\;\dagger}\lambda^a\vec{q}_{\alpha}\delta_{ab}\vec{q}_{\beta^\prime}^{\;\dagger}\lambda^b\vec{q}_{\beta}\;.
\]
Here, $\alpha$ and $\beta$ represent the color charges of the 
incoming quarks, $\alpha^\prime$ and $\beta^\prime$ of the 
outgoing quarks. Collecting the color parts in a color factor
\[
C^{\rm c}_{\alpha\alpha^\prime\beta\beta^\prime} =
\frac{3}{4}\sum_{a=1}^8\vec{q}_{\alpha^\prime}^{\;\dagger}\lambda^a\vec{q}_\alpha\,\vec{q}_{\beta^\prime}^{\;\dagger}\lambda^a\vec{q}_\beta
=
\frac{3}{4}\sum_{a=1}^8(\lambda^a)_{\alpha\alpha^\prime}\,(\lambda^a)_{\beta\beta^\prime}\;,
\]
one can calculate the net amplitude by summing over all possible
combinations of in- and outgoing colors. 
As there is evidence from lattice calculations that there is
no color transport over distances larger than 
$\lambda\approx 0.2\ldots 0.3\,\mathrm{fm}$, only the
commutating diagonal Gell-Mann matrices $\lambda_3$ and $\lambda_8$ 
from the Cartan subalgebra of $SU(3)_{\rm color}$ 
contribute over larger distances. In this Abelian approximation
the total color matrix for quark-quark interactions then is given by
\[
C^{\rm c}_{\alpha\beta} 
= \frac{3}{4}\sum_{a=3,8}(\lambda^a)_{\alpha\alpha}(\lambda^a)_{\beta\beta}
= \vec{w}_{\alpha}^{\sf T}\vec{w}_{\beta}\;,
\]
where
\[
\vec{w}_{\alpha} = \frac{\sqrt{3}}{2}\left(\begin{array}{c}
(\lambda^3)_{\alpha\alpha}\\ (\lambda^8)_{\alpha\alpha}\\ 
\end{array}\right),\quad \alpha =  1,2,3\;(R, G, B)
\]
are the normalized weight vectors corresponding to the three quark colors
in $(\lambda^3,\lambda^8)$ space.
Imposing a factor $-1$ at each antiquark vertex in color space 
yields the color matrix elements for the different color combinations 
as collected in table \ref{tab1}. They can easily be read off as the
scalar products of the weight vectors corresponding to the three colors or
anticolors, respectively. Positive values indicate attractive, negative
repulsive interactions. 

\Table{Color matrix elements of the 36 different
 elementary color combinations of the quarks. The matrix
elements can be obtained from the scalar products of the 
corresponding weight vectors}

\renewcommand{\arraystretch}{1.2}
\renewcommand{\arraycolsep}{5mm}
\begin{tabular}{l|cccccc}
\br
$C_{\rm c}^{\alpha\beta}$&$R$&$G$&$B$&$\overline{B}$&$\overline{G}$&$\overline{R}$\\
\hline
R&$-1$&$+\frac{1}{2}$&$+\frac{1}{2}$&$-\frac{1}{2}$&$-\frac{1}{2}$&$+1$\\
G&$+\frac{1}{2}$&$-1$&$+\frac{1}{2}$&$-\frac{1}{2}$&$+1$&$-\frac{1}{2}$\\
B&$+\frac{1}{2}$&$+\frac{1}{2}$&$-1$&$+1$&$-\frac{1}{2}$&$-\frac{1}{2}$\\
$\overline{\mathrm{B}}$&$-\frac{1}{2}$&$-\frac{1}{2}$&$+1$&$-1$&$+\frac{1}{2}$&$+\frac{1}{2}$\\
$\overline{\mathrm{G}}$&$-\frac{1}{2}$&$+1$&$-\frac{1}{2}$&$+\frac{1}{2}$&$-1$&$+\frac{1}{2}$\\
$\overline{\mathrm{R}}$&$+1$&$-\frac{1}{2}$&$-\frac{1}{2}$&$+\frac{1}{2}$&$+\frac{1}{2}$&$-1$\\
\br
\end{tabular}
\parbox{4.5cm}{\hfill\epsfig{figure=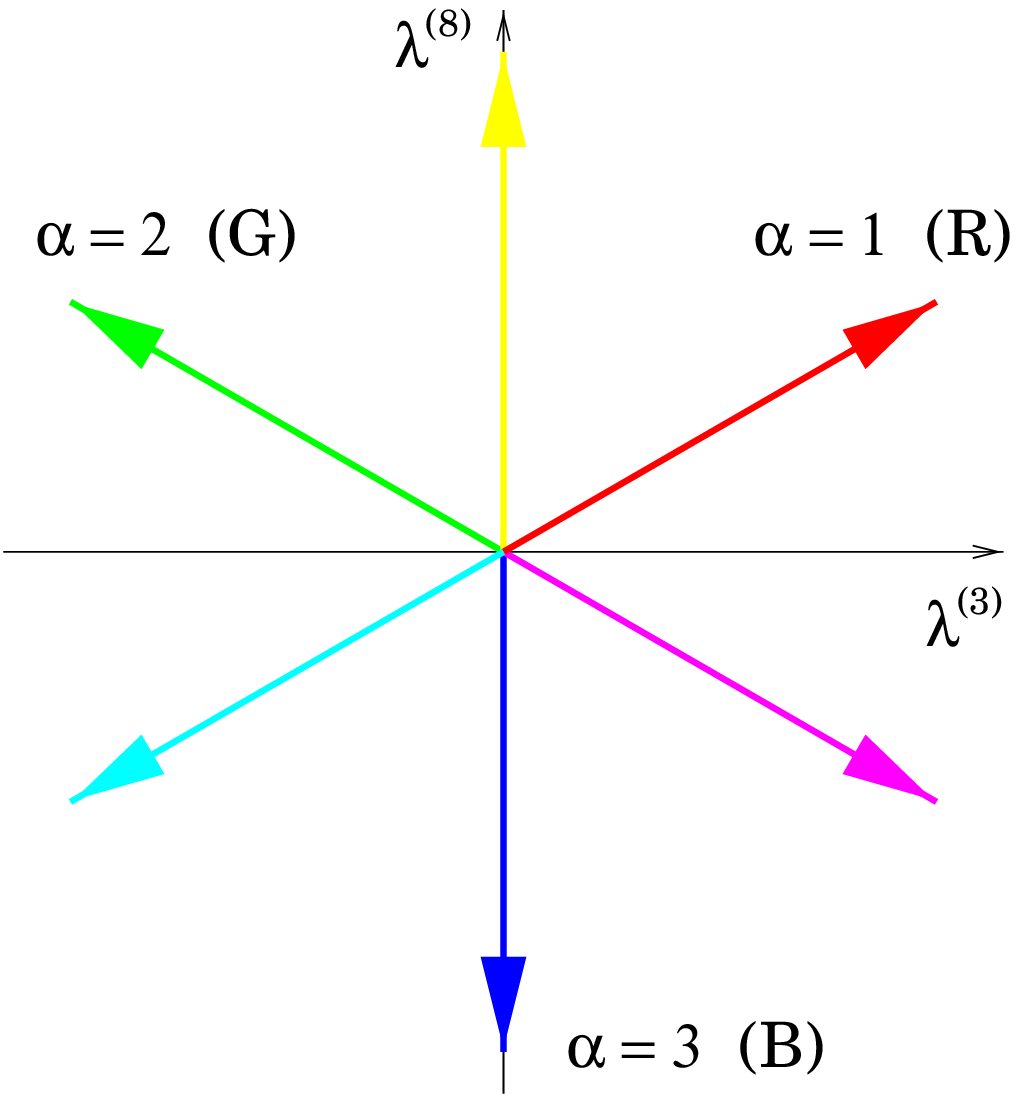,width=4cm}}
\label{tab1}
\endTable

Note that the relative strength of the color matrix elements
is rigorously enforced by  the requirement
of color neutrality of widely separated $q\overline{q}$ and $qqq$
states. 

The properties of the interacting quark gas turn out to be
independent from the selection of the shape of the potential 
at small distances, as far as the long distance term is defined properly. 
Therefore, we shall extend the linear potential to small
distances $r$ instead of using the color Coulomb potential 
at small $r$, which brings us in accordance to widely used 
phenomenological models for hadrons \cite{Regge:1959mz,Gribov:1968fc}. 

Regge trajectories yield values of $\kappa_0\approx 1.1\,{\rm GeV/fm}$,
while in the string model the string constant is found 
at $\kappa_0\approx 0.9\,{\rm GeV/fm}$.
However, these values were fitted to the properties of isolated
strings \cite{Sai91}. In a dense medium, quarks interact 
with all other color charges. This prohibits the confinement 
of the field lines into one single flux tube -- deconfinement is the
consequence.
Thus, free string constants $\kappa_0$ are not appropriate 
to calculate the properties of quark matter 
at high temperatures and densities as expected in heavy-ion 
collisions \cite{BIRO84}. 
In-medium effects, e.~g.\ interacting color fields, yield
an effective increase of the string tension (Casimir scaling) \cite{Fab98}. 
In the present model $\kappa$ effectively describes 
these in-medium effects. It will be treated as a free parameter of the
model and should  not be identified with the zero temperature value 
of free strings. 

Obviously, a sufficiently high density of color charge carriers
will lead to the screening of the the color interaction in the
dense medium, and thus color deconfinement results, even in
the simple semi-classical toy model presented here. We will discuss 
this below.

On the other hand, in a less dense and cooler system, all 
quarks will condense into clusters of two or three (anti-)particles 
with a total color charge in each cluster of zero. 
Note that higher quark numbers may also form  totally
color neutral states which appear to be bound. 
However, further propagation causes a separation 
into smaller likewise color neutral subclusters.
Therefore, we ultimately obtain bound states which correspond to
mesons or baryons.

\section{Hadronization}

\begin{figure}
\centerline{\epsfig{figure=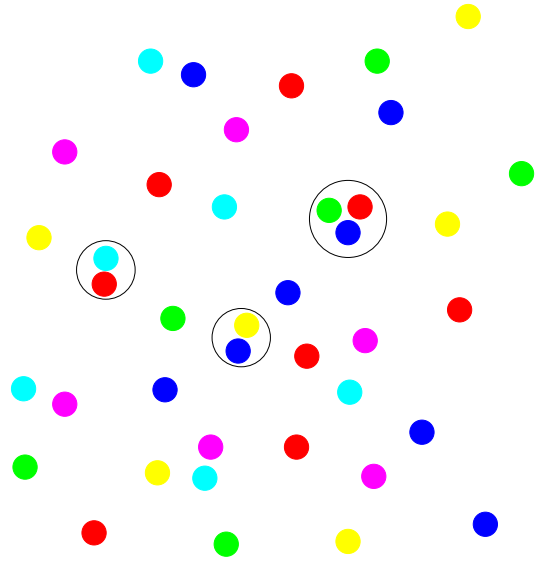,width=7cm}}
\caption{Hadronization of white quark clusters}
\label{hadro}
\end{figure}

It is now the second request to the model to define a criterion how to
map those bound quark states to hadrons. Such a mechanism
is essential as the Hamiltonian is not tuned to describe bound and
truly confined hadron states. Attempts have been made \cite{Bonasera:1999py}
to do so in a Vlasov approach. 
Here, we use the straight-forward requirement that the 
total color interaction from a pair 
(or a three particle state) of quarks with the remaining system
vanishes . 
Then, these $q\overline{q}$- and $qqq$-states do no longer
contribute to the color interaction of the quark gas (see figure \ref{hadro}). 
In the present model, this criterion of confinement -- which in a
numerical simulation of course  would never be fulfilled exactly -- has been
softened by introducing a lower bound for the remaining interaction $\kappa_\mathrm{min}$ between the cluster and the residual quark matter 
beyond which the cluster is declared to be frozen out \cite{Crawford:1982yz}. 
It is convenient to measure $\kappa_\mathrm{min}$ in units of the natural
scale of the model, $\kappa$.
\[
\left|\mathbf{F}_{\rm cluster}\right| = \left|\frac{1}{N_{\rm cluster}}\sum_{i\in{\rm
cluster}}\mathbf{F}_i\right| <\kappa_\mathrm{min} = F_{\rm cut}\cdot\kappa\;.
\]
Here,
 \[
\mathbf{F}_i = \sum_j\mathbf{F}_{ij} = -\sum_jC_{ij}\mathbf{\nabla}_jV(\left\vert\mathbf{r}_i-\mathbf{r}_j\right\vert)
\]
gives the total force of the system acting on particle $i$. 

If a bound quark state fulfills the hadronization criterion it will
be mapped to an appropriate hadronic state with identical quantum
numbers. Spin and isospin of the hadron is randomly chosen according
to the probabilities given by the Clebsch-Gordon coefficients.

The mass of the produced hadron is determined by energy
and momentum conservation. The total energy of the multi-quark state is
given by the expression
\[
E_\mathrm{H} =
\sum_{i\in\mathrm{cluster}}\biggl( E_i+\frac{1}{2}\sum_{{j\in\mathrm{cluster}\atop j\not= i}}C_{ij}V(\left\vert\mathbf{r}_i-\mathbf{r}_j\right\vert)\biggl)+\delta E\;.
\]
where $\delta E$ represents the residual energy which was set free due
to the field cut-off in the hadronization process and is of the order
$\delta E/E\lesssim 10^{-2}$. The momentum of the
hadron reads 
\[
\mathbf{P}_\mathrm{H} = \sum_{i\in\mathrm{cluster}}\,\mathbf{p}_i
\]
which yields a hadron mass of 
\[
M_\mathrm{H} = \sqrt{E_\mathrm{H}^2 - \mathbf{P}_\mathrm{H}^2}\;.
\]
Usually the obtained hadron masses will hardly fit to the tabulated
pole masses
of the known hadrons. Therefore, the quark clusters will preferably be
mapped to resonances with a broad mass distribution instead of sharply
peaked ground states. In case of multiple possible selections for
given quantum numbers we pick one randomly according to mass
distributions which are given by Breit-Wigner distributions
\[
f(M)\sim\frac{\Gamma^2}{(M-m_0)^2+(\Gamma/2)^2}\;.
\]
Here, $m_0$ and $\Gamma$ denote the peak mass and the total decay
width of the particle, respectively. To low masses, the distribution
is cut-off at a minimal mass to ensure hadronic decay according to the
experimentally known branching ratios. 
In the current version the model discriminates 29 mesonic and 36
baryonic states. 

\section{Thermodynamic properties of the interacting quarks gas}

In the present work, the properties of 
the interacting quark gas, i.\ e.\ of hot quark matter,
are studied in complete thermal equilibrium.
The system of interacting quarks is \emph{not} 
an ideal gas, but rather a strongly coupled quark fluid.
Therefore, the integration of the partition function 
cannot be carried out analytically. 
By adopting the Metropolis algorithm \cite{Met53}, 
an arbitrary number $N_{\rm rep}$ of 
$N$-particle phase space configurations 
can be generated. The latter configurations
are representations of the equilibrium state of the system. 
These representations are evaluated by imposing an initial
configuration $(\mathbf{x}_i^{(0)},\mathbf{p}_i^{(0)})\;(i=1..N)$ and then
repeatedly joggling all coordinates and momenta
\begin{eqnarray}
\mathbf{x}_k^{(r)}\;&\to\;\mathbf{x}_k^{(r+1)} = \mathbf{x}_k^{(r)}+\delta\mathbf{x}_k^{(r)}\;,\nonumber\\
\mathbf{p}_k^{(r)}\;&\to\;\mathbf{p}_k^{(r+1)} = \mathbf{p}_k^{(r)}+\delta\mathbf{p}_k^{(r)}\;.\nonumber
\end{eqnarray}
In each iteration, each displacement 
$(\delta\mathbf{x}_k^{(r)},\delta\mathbf{p}_k^{(r)})$ in phase space will cause a
change in total energy of the system
\[
\Delta E = E^{(r+1)}-E^{(r)}\;.
\]
According to the standard Metropolis algorithm,
if $\Delta E<0$, the new configuration is energetically more favorable
than the old one and will be accepted.
If, on the other hand, $\Delta E$ is positive, the new configuration
will be accepted with a probability $\exp(-\Delta E/T)$. This
allows for a statistical increase of the free energy of the system
driven by the ``temperature'' $T$. After a sufficient number of
iterations the system will enter a stationary state, where further
iteration will account for a thermal motion of the sample around 
the ground state. All configurations can
then be identified as representations of the thermalized state. 

Now, the ensemble average of any
thermodynamical variable $O$ can be approximated by the sum over 
those representations 
\[
\langle O \rangle = \frac{1}{N_{\rm rep}}\sum_{k=1}^{N_{\rm
rep}}O(\mathbf{x}_i^{(k)},\mathbf{p}_i^{(k)})\;,\quad i=1\ldots N\;.
\]
This enables us to calculate the energy density
\[
\epsilon = \frac{1}{V}\langle {\cal H} \rangle
\]
and -- by using the virial theorem --  the pressure of the interacting
quark gas
\[
P = \frac{1}{3V}\left\langle\sum_i\mathbf{p}_i\mathbf{v}_i+\sum_i\mathbf{r}_i\mathbf{\nabla}_iV\right\rangle\;.
\]
Here $\mathbf{v}_i = \mathbf{p}_i/E_i$ is the velocity of particle $i$.

In addition to the description of the quark phase we have to cope with
the hadronic sector. The produced hadrons are evaporated into the void.
The hadron pressure and temperature are assumed to be equal to 
the quark phase.

We will now assume a system of  $N$ quarks in a finite sphere of
volume $V$ with all color charges adding up to zero. 
This system is thermalized at a temperature $T$ according to the
previously discussed Metropolis method. A spherical system with a
radius of 4 fm contains about 400 quarks and antiquarks at a temperature
of 150 MeV.
If during the equilibration
process quarks form clusters that fulfill the above hadronization
criterion they are converted to color neutral hadrons which do no
longer interact due to color forces.

\subsection{Mixed phase and the equation of state}

\begin{figure}
\centerline{\epsfig{figure=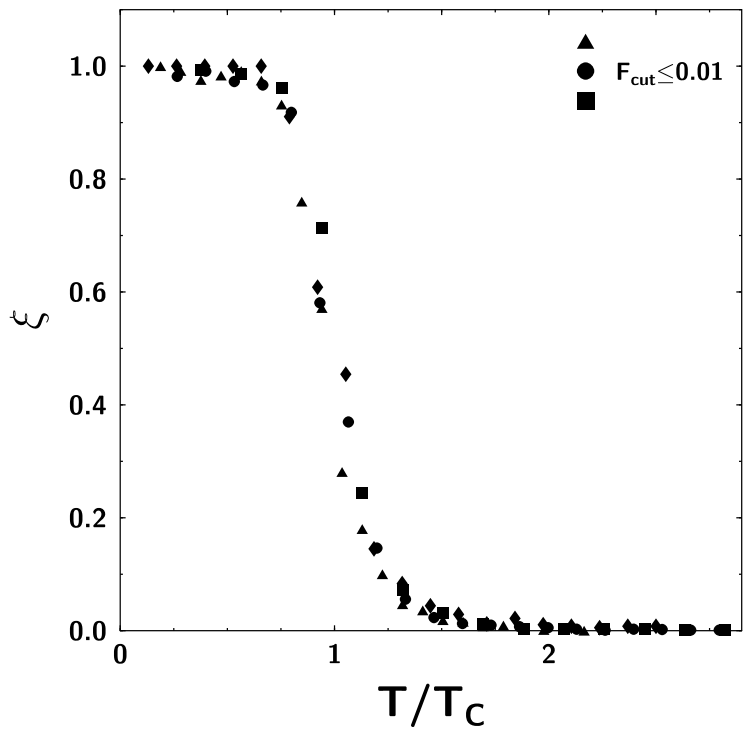,width=10cm}}
\caption{Hadron fraction as a function of temperature in a finite
quark blob for various sets of parameters $\kappa$ and
$F_\mathrm{cut}$. The temperature is measured in $T_\mathrm{C}$ which 
is defined for any set of parameters as the temperature of steepest 
descent with $T$. The distributions show perfect scaling behavior for 
$F_\mathrm{cut} < 0.01$ (black symbols).}
\label{xi_scale}
\end{figure}

The first important observable is the number of quarks which 
are hadronized from a given thermodynamic sample. For high
temperatures this quantity should converge to zero, while at $T\to 0$
all quarks should be hadronized due to confinement. The fraction $\xi =
N_\mathrm{h}/(N_\mathrm{h}+N_\mathrm{q})$ of hadrons
compared to the total particle number in the system therefore should
be $1$ in this limit. 
Figure \ref{xi_scale} depicts this hadron fraction $\xi$ as a function of
the temperature ($\mu=0$) measured in units of the ``critical
temperature'' $T_\mathrm{C}$, where the latter is defined as the temperature of
the steepest descent for each set of parameters $(\kappa,F_\mathrm{cut})$.
A rapid fall-off within $0.2\,T_\mathrm{C}$ from a hadron to a quark dominated
phase can be observed indicating the existence of a mixed phase during
the transition. In  case of a true  first order
phase transition in an infinite volume a sharp discontinuity of this quantity 
would be expected at $T_{\rm C}$ \cite{SPIELES97a}. 

\begin{figure}
\centerline{\epsfig{figure=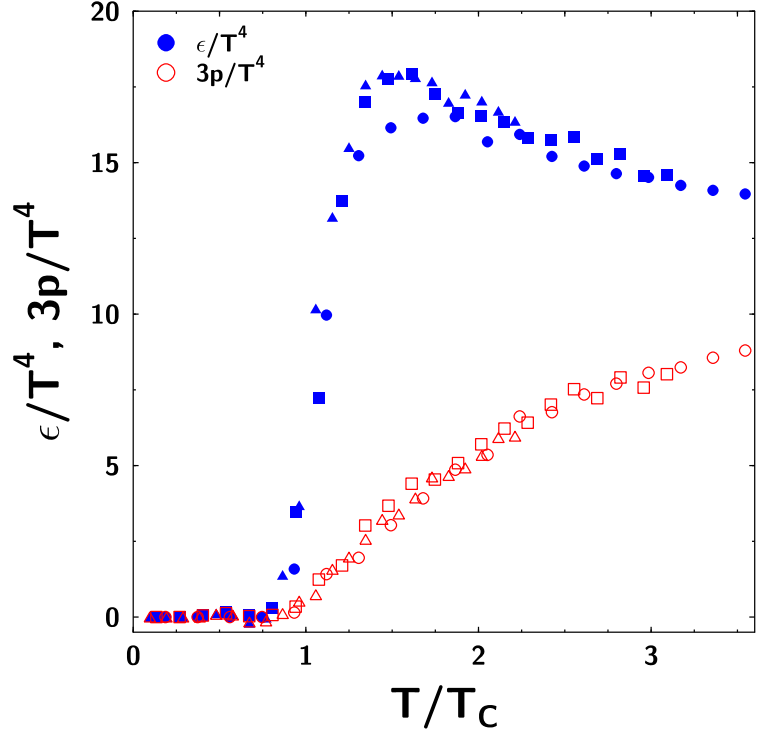,width=10cm}}
\caption{Energy density and pressure of the quark phase 
as a function of temperature for various sets of parameters $\kappa$ and $F_\mathrm{cut}$. }
\label{3inf}
\end{figure}

A similar continuous transition can be observed in the
energy dependence of the quark phase as plotted in figure
\ref{3inf}. Here, the energy density $\epsilon$ and the pressure $p$
are  divided by $T^4$ and are given for various values of $\kappa$ and
$F_{\rm cut}$ as discussed above. The pressure has been multiplied by
a factor of $3$. Lattice calculations reveal a similar transition, 
slightly smoothed for energy density and  pressure.
\cite{Kar95}. However, our microscopic finite size simulation exhibits 
an even broader crossover. It is worth to note that the absolute values 
of lattice calculations for very high temperatures may not be compared 
to our results as we neglect the contributions of hard gluons. 

\begin{figure}
\centerline{\epsfig{figure=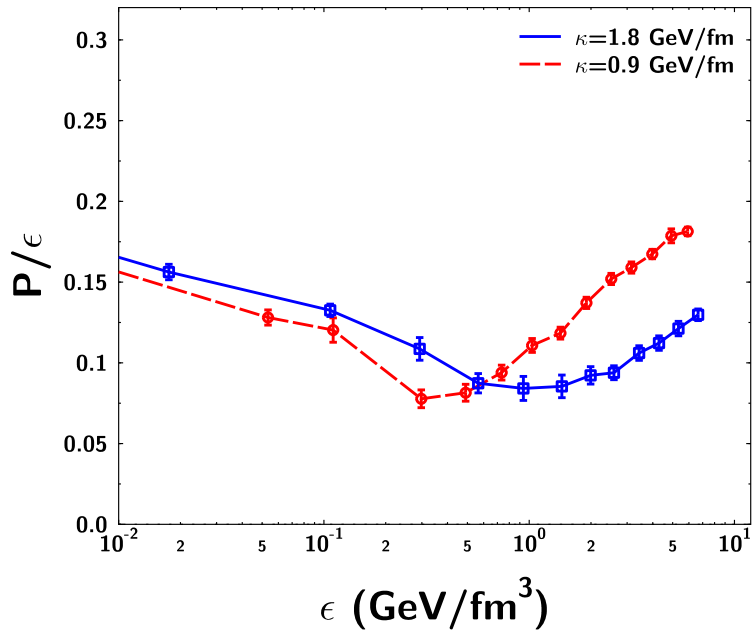,width=10cm}}
\caption{Equation of state for $\kappa=\kappa_0$ (dashed line) and
$\kappa=2\kappa_0$ (solid line). A softening of the EOS around 
$\epsilon=1 \mathrm{GeV/fm}^3$ is revealed. }
\label{eos}
\end{figure}

A functional form of thermodynamic quantities similar to one found here
has been parameterized \cite{Ris96b,Asa97} in order to model the {\em assumed} 
smooth crossover transition  and to  study the  physical consequences. 
In accordance to those investigations our microscopic model also reveals a
minimum of the equation of state in the phase transition
region (see figure \ref{eos}). However, compared to the case of
infinite matter this dip is less pronounced.

The plots in fig.\ \ref{xi_scale} and \ref{3inf} both reveal a perfect 
scaling behavior for  $F_\mathrm{cut}\lesssim 0.01$.  
This imposes a natural range for the seemingly
completely arbitrary parameter $F_\mathrm{cut}$ which could a priori
not be connected to any physical observable. 

Despite the conformity in shape, the absolute scale of
$T_\mathrm{C}$ is strongly affected by the particular choice of these 
parameters. A reduction of $F_{\rm cut}$ to zero will ultimately 
shift $T_{\rm C}\to 0$, since hadronization is completely suppressed 
in this limit. On the other hand, an increase  of the string tension, 
$\kappa$, gives rise to an increasing critical temperature $T_{\rm C}$, 
revealing a scaling dependence of the form 
$T_{\rm C}\sim \sqrt{F_{\rm cut}\cdot\kappa}$.
This can be directly understood from the hadronization mechanism: 
If any colorless quark cluster ($q\overline{q}$, $qqq$ or, 
in principle, any multi-quark state) separates from the remaining quarks, 
we shall always obtain a finite remaining color field between 
the two quark samples whose strength $\kappa$ is reduced (screened) 
compared to the vacuum value. 
The quark clusters are now assumed to separate sufficiently slowly so
that the mediating color field lines can be considered to confine 
to an equilibrated flux tube which approximately fulfills the
presumptions of a cylindrical MIT bag. In the bag model the field strength
$\kappa$ is connected to the bag constant $B$ according to
$\kappa \sim \sqrt{B}$ \cite{Sai91}. On the other hand,
the bag pressure for an ideal quark-gluon gas raises as $B\sim T_{\rm C}^4$. 
-- therefore one immediately obtains $T_C\sim\sqrt{\kappa}$.
In our model, the separating cluster is declared a ``hadron'' 
if the remaining force drops below the cut-off $F_{\rm cut}\cdot\kappa$. 
Applying the flux tube picture as derived form the MIT bag model 
then yields the previously found dependency for the critical 
temperature $T_{\rm C}\sim \sqrt{F_{\rm cut}\cdot\kappa}$.

\begin{figure}
\centerline{\epsfig{figure=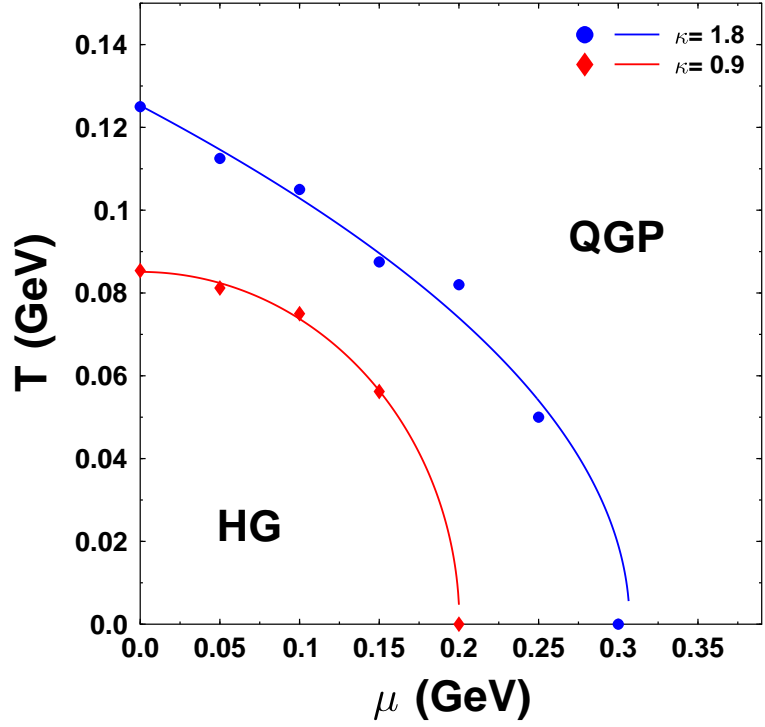,width=10cm}}
\caption{Phase diagram  in the $T$-$\mu$ plane for $\kappa/\kappa_0 =
1,2$. The lines are fits to the calculation. }
\label{mu}
\end{figure}

While the lattice results predict a critical temperature
of $T_{\rm C}\approx 150\,{\rm MeV}$ \cite{Kar95}, the natural choices
$\kappa=\kappa_0\approx 1\,{\rm GeV/fm}$ and 
$F_{\rm cut}=0.01$ give a much lower value $T_{\rm C}\approx 90\,{\rm MeV}$.
As the value for $F_{\rm cut}$ is at the upper bound 
of the scaling domain, the critical temperature may only 
be enhanced by increasing the field strength $\kappa>\kappa_0$.  
The impact of a variation 
of $\kappa$ on the thermodynamical properties 
is visualized in figure \ref{mu}. 
In this plot, we further extend the investigations 
to finite $\mu$ and calculate the phase diagram for 
various $\kappa$. First attempts to apply 
lattice QCD to finite densities \cite{Eng99} seem to support our
findings. It is obvious that for high $\kappa$ the curvature 
of the lines appears smaller as compared to the curves 
usually extracted from the MIT bag model.
To approach the lattice results of $T_\mathrm{C} \approx 140$ MeV
for $\mu\to 0$, values $\kappa\gtrsim 2 \kappa_0$ are required.
For $T\to 0$ the chemical potential than is about $350\, \mathrm{MeV}$ 
so that normal nuclear matter ($\mu \approx 300$ MeV) 
is safely within the hadronic region. This is in perfect 
agreement with the above discussion and reflects 
the finite density character of the quark phase. 

The high value found for $\kappa$ should not come as a surprise: 
As discussed before, because of in-medium effects an increased string
constant (compared to the free value) should be expected.

However, this high value of $\kappa$ contradicts to the two-particle limit 
(one quark and one antiquark). The model then turns into the common 
string model which determines the color field strength to $\kappa = \kappa_0$. 
In principle, this suggests the introduction of 
a density and temperature dependence of the string
constant. A  more qualitative view on the properties of the 
interacting quark gas shall be provided. 
However, it should be noted that the hadronization of the QGP is
mainly based on quark rearrangements within the blob compared to string
fragmentation on the surface of the plasma. This also supports the
assumption of particle number conservation in the hadronization process.
Hence, we fix the string constant to a medium value of $\kappa=2\kappa_0$.

\begin{figure}
\centerline{\epsfig{figure=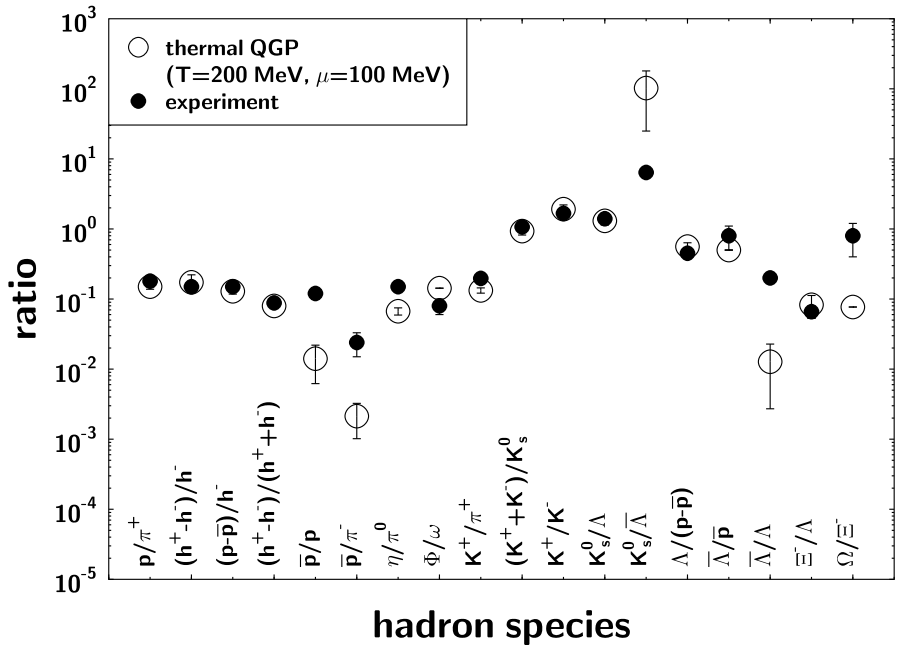,width=12cm}}
\caption{Final state hadron ratios from thermal qMD calculations (open
circles) compared to S+Au data at 200 $A$GeV (full circles, taken from
\cite{BrMun96})}
\label{ratios}
\end{figure}

The hadronization of the thermalized quark system yields hadron ratios
which can be compared at mid-rapidity to those measured in CERN-SPS 
experiments (see compilation in \cite{BrMun96}). Fig.~\ref{ratios} shows
the comparison to the outcome in a S+Au collision, assuming a
thermal fireball as hadron source.
We find a very good agreement in all $MM$, $MB$ and $BB$ ratios, while
the antibaryons seem to be clearly under-predicted. 
Particle ratios, however, proved not to be a very sensitive observable
to test the quality of theoretical models. Fits of a pure hadron gas
\cite{BrMun96} proved to describe data with a comparable precision as 
other thermal or hydrodynamical approaches including a QGP phase
transition or several microscopic simulations as UrQMD \cite{Bass:1997xw}.
However, the analysis of event-by-event fluctuations \cite{Bleicher:1998wu}
and of the dynamical properties of the system may yield new insight.

\subsection{Dissociation of a quark blob}

\begin{figure}
\centerline{\epsfig{figure=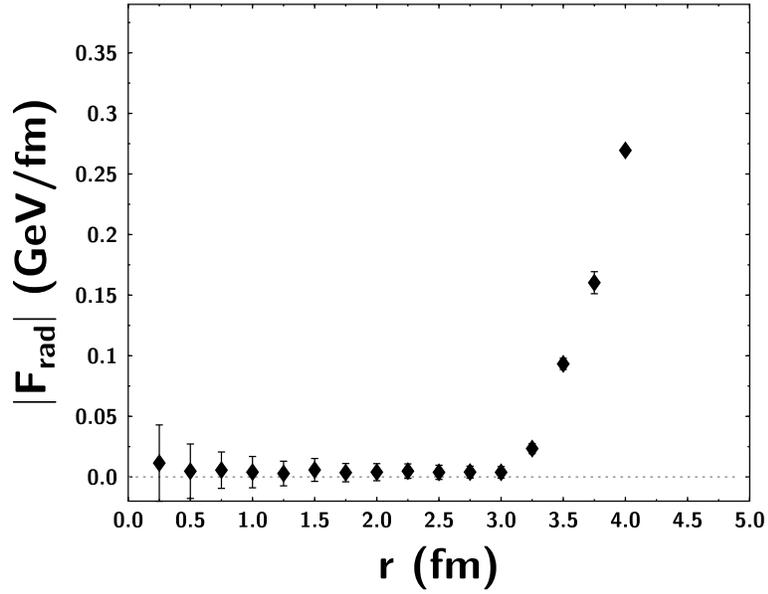,width=10cm}}
\caption{Averaged radial force $\left\vert\sum\mathbf{F}\hat{\mathbf{r}}\right\vert$
acting on a quark at a distance $r$ 
within a quark blob of radius $R=4 \mathrm{fm}$. In the center the quarks
are approximately free. Near the surface they are strongly pulled back
into the sphere.}
\label{forces}
\end{figure}

\begin{figure}
\centerline{\epsfig{figure=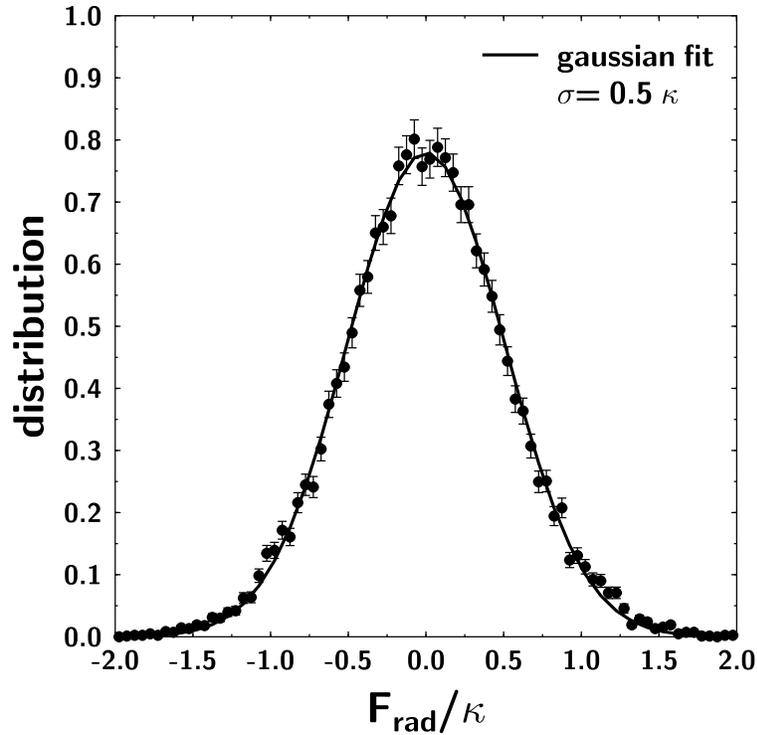,width=10cm}}
\caption{Fluctuations of net  radial forces
$F_\mathrm{rad}=\sum\mathbf{F}\hat{\mathbf{r}}$ acting on a central quark
($r<1 \mathrm{fm}$).}
\label{flucts}
\end{figure}

All results from the last paragraph presume complete equilibration of a
finite canonical ensemble which is defined by  all possible
microscopic representations at a time. One particular representation
will always contain 
fluctuations which may cause the properties of the single
representation to deviate strongly from the collective behavior
\cite{Ble98}. This effect is emphasized in  figure \ref{forces} where
the average radial force 
\[
F_\mathrm{rad}(r) = \left\langle\sum_i\mathbf{F}_i\,\hat{\mathbf{r}}_i\right\rangle
\]
acting on a quark at a distance $r$ from the origin of a 
spherical thermalized quark blob ($R=4\, \mathrm{fm}$, 
$T=200\, \mathrm{MeV}$, $\mu=100\, \mathrm{MeV}$) is plotted. 
In the ensemble average the quarks in the center of the quark matter 
do not feel any net interaction: color is screened. 
A net interaction within approximately $1 \mathrm{fm}$ from the surface
traps the color charges confined within the blob. 
Moreover, this result is \emph{independent} of the
particular shape of the interaction potential as long as it fulfills
the symmetry requirements concerning the color charges as given in
table \ref{tab1}. Then, within the center of the quark phase all
contributions from the potential cancel exactly to zero if the spatial
distribution is sufficiently  homogeneous.
However, this statement holds only for a large number of quark
samples. In one single microscopic representation one can find large
fluctuations of the net color force on each quark. This is
pointed out in figure \ref{flucts}, where the distribution function of
the radial color forces $\mathbf{F}_i\,\hat{\mathbf{r}}_i$ acting on quarks
\emph{in the center} of the blob is plotted. We obtain an almost
perfect Gaussian distribution with a standard deviation
$\sigma=0.5\kappa$ indicating huge fluctuations. Therefore strong
inhomogeneities in one single event are to be expected. 
Hence, the microscopic system will not behave like an ideal
quark gas. Due to the color interactions we do not expect that 
the quark system does expand hydrodynamically, smoothly reducing 
temperature and density. Instead, these results indicate that 
during the expansion the quark phase will rupture, and hadrons will
condense both from its surface as well as from its interior.

\section{Conclusion}

We have presented a microscopic description of the deconfinement
phase transition by means of a semiclassical interacting quark gas 
supplemented with a dynamical hadronization criterion. 
The color interaction potential has been motivated from phenomenological QCD 
in the abelian approximation. A smooth crossover was found, 
comparable to recent lattice results. The phase diagram
for finite $\mu$ has been calculated. Particle ratios have been compared
to experimental results yielding a reasonable agreement.
An event-by-event analysis revealed strong fluctuations which 
initially drive the dissociation process.

\section*{Acknowledgments}
This work is supported by  GSI, BMBF, DFG, Graduiertenkolleg
Theoretische und Experimentelle Schwerionenphysik, and
the Josef Buchmann Foundation.

\newpage

\end{document}